\documentclass{aa}  
\usepackage{graphicx}
\usepackage{txfonts}
\usepackage{color}
\usepackage{booktabs}   
\usepackage{multirow}   
\usepackage{multicol}   
\usepackage{booktabs}
\usepackage{multirow}
\usepackage{makecell}

\newcommand{\HI}{{\sc H\,i }}
\newcommand{\rhi}{$R_\mathrm{HI}\,$}

\begin{document} 

   \title{Dark matter in ALFALFA galaxies}

   \subtitle{Investigating galaxy-halo connection}

   \author{M. Yang\inst{1}\fnmsep\thanks{\email{myang@shao.ac.cn}},
          L. Zhu\inst{1}\fnmsep\thanks{\email{lzhu@shao.ac.cn}},
          N. Yu\inst{2,3},
          Y. Lei\inst{1},
          R. Cai\inst{1},
          J. Wang\inst{3} \and
          Z. Zheng\inst{3}
          }          

   \institute{Shanghai Astronomical Observatory, Chinese Academy of Sciences, 80 Nandan Road, Shanghai 200030, China\and
             Max Planck Institute for Radio Astronomy, Auf dem Hügel 69, 53121 Bonn, Germany \and
             National Astronomical Observatories, Chinese Academy of Sciences, A20 Datun Road, Chaoyang District, Beĳing, 100101, China 
             }

   \date{Received 17 June 2025; accepted 15 December 2025}

  \abstract
   {}
   {This paper aims to investigate the galaxy-halo connection using a large sample of individual galaxies with \HI integrated spectra. We determine their dark matter content by applying a dynamical method based on \HI line widths measured with the curve-of-growth technique, together with inclination corrections inferred from optical images.}
   {
   We build a sample of 2453 gas-rich predominantly late-type galaxies spanning a stellar mass range of $10^{8.7}M_\odot$ to $10^{11.4}M_\odot$ by matching them one-to-one with their counterparts from the ALFALFA survey and the TNG100 simulation, ensuring a direct match of stellar mass and \HI radius. We generate mock images and mock \HI integrated spectra for TNG100 galaxies, and apply the same dynamical method to both ALFALFA and TNG100 mock galaxies to infer their dark matter masses.
   }
   {   
   Across all stellar mass bins, ALFALFA galaxies exhibit lower median dark matter masses than the mock TNG100 simulation results. In each bin, this offset is driven by a tail of galaxies with comparatively low dark matter content, which becomes more prominent toward higher stellar masses. In the highest mass bin ($M_* > 10^{11} M_\odot$), late-type ALFALFA galaxies show a median dark matter mass that is 23\% lower than that of their counterparts in the TNG100 dark-matter-only simulation, with 32\% of ALFALFA galaxies having $M_\mathrm{DM}(<R_\mathrm{HI})<10^{11.5} M_\odot$, compared to 17\% in the mock TNG100 sample. These results suggest that a larger fraction of massive late-type galaxies reside in relatively less massive dark matter haloes than predicted by the TNG100 simulation.
   }
   {}


   \maketitle


\section{Introduction}
In cosmological models, galaxies form in the central potential well of dark matter haloes. The density profiles of dark matter haloes are determined by the attributes of dark matter particles: Without the baryonic effect, cold dark matter forms central cusps~\citep[e.g.][]{1996ApJ...462..563N,2008MNRAS.391.1685S} while self-interacting warm dark matter forms flatter cores~\citep[e.g.][]{2016PhRvL.116d1302K,2017PhRvL.119k1102K}. 

The connection between galaxies and their dark matter haloes also affects the amount of dark matter in galaxies. The stellar-to-halo mass relation (SHMR) and its scatter have been measured in various approaches, such as abundance matching~\citep[e.g.][]{2010ApJ...710..903M, 2010ApJ...717..379B, 2013ApJ...771...30R}, galaxy clustering~\citep[e.g.][]{2007ApJ...667..760Z, 2009ApJ...695..900Y, 2015MNRAS.454.1161Z, 2017ApJ...839..121T}, satellite kinematics~\citep[e.g.][]{2007ApJ...654..153C, 2011MNRAS.410..210M, 2013MNRAS.428.2407W}, galaxy-galaxy lensing~\citep[e.g.][]{2006MNRAS.368..715M, 2016MNRAS.457.3200M, 2020MNRAS.492.3685H, 2021A&A...653A..82B}, and empirical models~\citep[e.g.][]{2015ApJ...799..130R, 2019MNRAS.488.3143B}. The halo mass is typically defined as $M_{200}$, the mass enclosed within a sphere of mean density 200 times the critical density of the Universe. These findings indicate that massive galaxies reside in massive dark matter haloes, and the scatter of SHMR is influenced by galaxy colours or morphologies, with dark matter haloes of early-type galaxies being more massive than those of late-type galaxies with equivalent stellar mass, especially at the high-mass end. However, most of these results are obtained by stacking the weak or undetectable signals of individual galaxies from large samples. 

The direct determinations of SHMR require dark matter measurements in individual galaxies. Large samples of stellar kinematic data from spatially resolved integral field unit (IFU) spectrographs have been used to build dark matter distributions in galaxies~\citep{2013MNRAS.432.1709C,2022ApJ...930..153S,2023MNRAS.522.6326Z}. However, the limited field-of-view of these measurements covering galaxy bright central regions can only measure dark matter distributions within a few effective radii of galaxies with significant uncertainties, which might only provide constraints on baryonic feedback during galaxy formation. Extended IFU and long-slit spectrographs~\citep[e.g.][]{2010ApJ...716..370F,2016MNRAS.460.3029B,2023A&A...675A.143C} and other tracers such as \HI rotation curves~\citep[e.g.][]{2008AJ....136.2648D,2011AJ....141..193O,2020ApJS..247...31L,2020ApJ...905...28H,2020MNRAS.491.4221Y,2022MNRAS.512.1012C}, planetary nebulae and globular clusters~\citep[e.g.][]{2016MNRAS.460.3838A,2017MNRAS.468.3949A,2024A&A...685A.132D}, hot interstellar X-ray gas~\citep[e.g.][]{2019ApJ...887..259H,2020ApJ...905...28H}, and stellar streams~\citep{2022ApJ...941...19P,2023ApJ...954..195N} can probe dark matter on the outskirts of extragalactic galaxies, yet there is no large sample because of the expensive observations. Until now, direct determinations of SHMR have been reported only in a handful of studies. These include analyses of central galaxies in galaxy groups and clusters~\citep{2018AstL...44....8K, 2019A&A...631A.175E}, as well as dynamical studies of individual galaxies using \HI kinematics~\citep{2019A&A...626A..56P, 2022MNRAS.514.3329M, 2023MNRAS.518.6340D, 2023MNRAS.524.6213B, 2024ApJ...971...69D} or globular cluster kinematics~\citep{2021A&A...649A.119P}.

For a better understanding of the connection between galaxies and their halos, it is essential to establish dark matter measurements at extended radii for large samples of individual galaxies. Compared to other time-consuming tracers, \HI integrated spectra provide a cost-effective method to estimate dark matter content. For gas in dynamic equilibrium, using the rotational velocity derived from the \HI line width in conjunction with the galaxy sizes is the most efficient method to estimate galaxy dynamical mass~\citep[e.g.][]{1980A&A....81..371C,1997A&A...324..877B,2018MNRAS.478.1611K} and even dark matter halo mass $M_{200}$~\citep{2020ApJ...898..102Y}. The Arecibo Legacy Fast ALFA survey~\citep[ALFALFA;][]{2018ApJ...861...49H} identified approximately $31500$ extragalactic \HI emissions within the redshift of about $0.06$, and its spectra have been applied to provide measurements of the dynamical mass within the \HI radius of galaxies~\citep{2022ApJS..261...21Y}, which can be used to further deduce the dark matter mass within the \HI radius. However, we cannot obtain the virial mass of dark matter haloes directly from such measurements; therefore, we need to compare these measurements with the dark matter mass within the \HI radius for galaxies in cosmological simulations for a meaningful interpretation, as earlier studies have contrasted the measured SHMR with multiple simulations~\citep[e.g.][]{2020MNRAS.499.3578C,2021NatAs...5.1069C,2025ApJ...980..233W} and compared the dark matter content within large radii of observations and simulations~\citep[e.g.][]{2017MNRAS.468.3949A,2018MNRAS.481.1950L,2020ApJ...905...28H,2024MNRAS.528.5295Y}, highlighting that observations show larger scatter in the SHMR, with a higher portion of galaxies exhibiting lower dark matter fractions compared to simulations.

This paper aims to explore the galaxy-halo connection by investigating the dark matter mass within the \HI radius for galaxies in the ALFALFA survey and the IllustrisTNG simulation. The structure is as follows: In Section~\ref{sec:method}, we introduce the method for measuring the dark matter content in galaxies; Section~\ref{sec:dands} covers data and sample description. In Section~\ref{sec:results}, the method is validated by applying it to mock galaxies in the simulation and comparing those results with true values; we present the dark matter mass of the ALFALFA galaxies alongside comparisons with the simulation. We discuss our main result in Section~\ref{sec:discussion}. The paper is summarised with Section~\ref{sec:summary}.

\section{Methods}
\label{sec:method}
The technique for determining galaxy dynamical mass using the \HI line width from an \HI integrated spectrum via the curve-of-growth method and the inclination from an optical image is detailed in~\citet{2020ApJ...898..102Y,2022ApJS..261...21Y}; we provide a brief overview here.

We measure the observational axis ratio $q$ with the optical image. Assuming an intrinsic disc thickness $q_0$ varying with the stellar mass $M_*$ according to the relation presented in~\citet{2010MNRAS.406L..65S}, the galaxy inclination $i$ is then
\begin{equation}
    \cos^2{i}=\frac{q^2-q_0^2}{1-q_0^2}.
\end{equation}
The \HI integrated spectrum is used to measure the \HI line width $V_{85}$, which is the velocity width that captures 85\% of the total flux. A rotation velocity $V_\mathrm{85,c}$ is then derived by correcting for the inclination (observational effects are omitted in this equation but included in measurements)
\begin{equation}
    V_\mathrm{85,c} = \frac{V_{85}}{2\sin{i}}.
\end{equation}
The circular velocity $V_\mathrm{c}$ at the \HI radius \rhi is further derived by this scaling relation (see Equation 9 of \citet{2020ApJ...898..102Y} for details),
\begin{equation}
    V_\mathrm{c} = (0.94\pm0.02)V_\mathrm{85,c}+(13.33\pm3.31){\rm km/s},
\end{equation}
The \HI radius \rhi is defined by where the \HI column density falls to $1\,M_\odot\,\mathrm{pc}^{-2}$. The \HI mass $M_\mathrm{HI}$ measured from the total flux of the \HI integrated spectrum has a tight empirical correlation with \rhi in isolated gas-rich galaxies, such as the calibration in~\citet{2016MNRAS.460.2143W},
\begin{equation}
    \log \left( \frac{R_\mathrm{HI}}{\rm kpc} \right) = (0.51\pm0.00) \log \left( \frac{M_\mathrm{HI}}{M_\odot}\right)-(3.59\pm 0.01).
    \label{equ:mhi-rhi}
\end{equation}
The dynamical mass within \rhi is then derived by
\begin{equation}
    M_\mathrm{dyn}(<R_\mathrm{HI}) = \frac{V_\mathrm{c}^2 R_\mathrm{HI}}{G} = 2.31\times10^5 M_\odot \left( \frac{V_\mathrm{c}}{\rm km/s}\right)^2\left( \frac{R_\mathrm{HI}}{\rm kpc} \right).
\end{equation}

The baryonic mass within $R_\mathrm{HI}$ includes the stellar, atomic, and molecular gas components. The stellar mass within \rhi is taken as the stellar mass of the galaxy $M_*$, since \rhi typically extends beyond the stellar disc in gas-rich galaxies. The atomic gas mass is $1.33\,M_\mathrm{HI}(<R_\mathrm{HI})$ to account for helium, where $M_\mathrm{HI}(<R_\mathrm{HI})$ is obtained by integrating the total HI mass $M_\mathrm{HI}$ assuming the normalised surface density profile of~\citet{2014MNRAS.441.2159W}:
\begin{equation}
    \Sigma_\mathrm{HI}(r) = \frac{\Sigma_0 \exp{(-r/r_s)}}{1+\alpha\exp{(-r/r_c)}},
\end{equation}
where $r=R/R_\mathrm{HI}$, and we use $\alpha=39.878$, $r_s=0.188$, $r_c=0.183$, with $\Sigma_0$ treated as a variable central density to ensure the integral equals $M_\mathrm{HI}$. The molecular gas is estimated from the star formation rate as follows~\citet{2024A&A...687A.244H}:
\begin{equation}
    \log \left( \frac{M_\mathrm{mol}}{M_\odot} \right) = (1.02\pm0.04) \log \left( \frac{\mathrm{SFR}}{M_\odot\,\mathrm{yr}^{-1}} \right) + (8.90\pm0.02) \pm 0.29.
\end{equation}
The baryonic mass within \rhi is then
\begin{equation}
    M_\mathrm{bary}(<R_\mathrm{HI}) = M_* + 1.33\,M_\mathrm{HI}(<R_\mathrm{HI}) + M_\mathrm{mol}.
    \label{equ:bary}
\end{equation}
We finally derive the enclosed dark matter mass within \rhi by 
\begin{equation}
    M_\mathrm{DM}(<R_\mathrm{HI}) = M_\mathrm{dyn}(<R_\mathrm{HI}) - M_\mathrm{bary}(<R_\mathrm{HI}).
\end{equation}

\section{Data and sample}
\label{sec:dands}
Our goal is to understand the SHMR of galaxies. However, the available observational data — optical images and \HI integrated spectra — provide only a single measurement of the enclosed dark matter mass within the \HI radius $M_\mathrm{DM}(<R_\mathrm{HI})$ for each galaxy, with $R_\mathrm{HI}$ varying from galaxy to galaxy. To interpret the observed $M_\mathrm{DM}(<R_\mathrm{HI})$ in the context of the SHMR, we compare these measurements with mock observations from the TNG100 simulation. By applying the same measurement procedure to the simulated galaxies, we can also quantify the systematic biases introduced by our method.

In this section, we describe the observational datasets, introduce the simulated galaxies, and outline the procedure for generating mock observations and performing the one-to-one matching between the observational and simulated samples.

\subsection{Observational data}
\label{sec:obs}
We choose our parent sample from the catalogue presented in~\citet{2022ApJS..261...21Y}, which is the overlap of the ALFALFA survey, the 16th data release of the Sloan Digital Sky Surveys~\citep[SDSS DR16][]{2020ApJS..249....3A} and the second version of the GALEX-SDSS-WISE Legacy Catalogue with the deepest photometry~\citep[GSWLC-X2;][]{2016ApJS..227....2S,2018ApJ...859...11S}. This catalogue measures the \HI mass, the \HI radius \rhi and the circular velocity $V_\mathrm{c}$ from the \HI integrated spectra of the ALFALFA survey to provide the dynamical mass within \rhi. It also collects optical and spectroscopic parameters including the $r$-band axis ratio $q$ from SDSS DR16 and the stellar mass $M_*$ and the star formation rate (SFR) measured with UV/optical/IR SED ﬁtting from GSWLC-X2. 

We also require that these parameters reported in the catalogue for the parent sample meet the following criteria:
\begin{itemize}
    \item The difference between the recession velocity of the optical spectra and the central velocity of the \HI spectra is required to be less than $30\,\mathrm{km/s}$. This threshold is motivated by the observed distribution of velocity offsets: most galaxies lie within this range, whereas larger differences (up to $\sim 200\,\mathrm{km\,s^{-1}}$) may indicate misidentifications, unsettled discs, strong misalignments, or external disturbances such as ram-pressure stripping. In the absence of resolved velocity fields, such systems cannot be reliably identified, and the cut therefore serves as a conservative safeguard to select galaxies where the \HI gas is likely in dynamical equilibrium.

    \item The signal-to-noise ratio of the \HI integrated spectrum is required to exceed $3.0$ to ensure a robust detection.

    \item The \HI concentration parameter $C_V$ is required to be below $4.0$. This parameter, defined as the ratio between the line widths $V_{85}$ and $V_{25}$ ~\citep{2022ApJS..261...21Y}, characterises the degree of concentration of the integrated profile. Values below this threshold correspond to broader, double-horned spectra, whereas higher values indicate more centrally concentrated, single-peaked profiles. Based on inspection of our sample, the adopted cut effectively selects galaxies with double-horned spectra.

    \item The inclination derived from the optical axis ratio is required to satisfy $\sin(i)\geq 0.7$ ($i \gtrsim 45^\circ$). This cut ensures that galaxies are nearly edge-on, thereby reducing uncertainties in the deprojection of velocities and enabling accurate dynamical mass estimates (see Appendix~\ref{appendix: tng100} for details).
\end{itemize}
We finally conduct a visual inspection of galaxy images to exclude irregular and face-on bar galaxies. This results in a parent sample of 4844 galaxies. We show the normalised \HI spectra of the parent sample in Figure~\ref{fig:ALFALFA_sample} (see Appendix~\ref{appendix: alfa_sample}) to illustrate the typical double-horned line profiles of rotation-supported discs. We obtain the baryonic mass within \rhi and derive the enclosed dark matter mass and dark matter fraction within \rhi for this parent sample. 

\subsection{Simulated data}
The IllustrisTNG project consists of a suite of state-of-the-art hydrodynamic cosmological simulations in large cosmological volumes. We utilise the `TNG100-1' simulation, which has a cubic volume with a side length of $110.7\,\mathrm{Mpc}$, a baryonic mass resolution of $1.4\times10^6 M_\odot$, and a dark matter resolution of $7.5\times10^6 M_\odot$. We use the snapshot 99 which corresponds to $z = 0$. We also utilise the dark-matter-only simulation called `TNG100-1-Dark', which has a dark matter resolution of $8.9\times10^6 M_\odot$.

We generate mock SDSS $r$-band images for TNG100 galaxies using the method described in~\citet{2025A&A...695A.177C} by projecting each galaxy onto a two-dimensional plane in a random direction. From these mock images, we measure the effective radii $R_\mathrm{e}$ and the axis-ratios $q$. We define the stellar mass $M_*$ as the total mass of stellar particles contained within $5 R_\mathrm{e}$ to estimate the intrinsic disc thickness $q_0$. The inclination $i$ is subsequently derived from the axis-ratio and the intrinsic disc thickness.

The \HI spatial distribution for TNG100 galaxies is generated with the \texttt{Hdecompose} package\footnote{https://github.com/kyleaoman/Hdecompose}. This implementation calculates the neutral gas fraction of each gas particle based on the multiphase ISM model detailed in~\citet{2003MNRAS.339..289S}. We utilise the `volumetric' method as outlined in~\citet{2018ApJS..238...33D} and originally introduced by~\citet{2015MNRAS.452.3815L} to calculate the molecular gas fraction, employing the $\rm H_2$ model by~\citet{2011ApJ...728...88G}. 

We describe the \HI emission from each gas particle with a Gaussian profile centred on the velocity of the particle. We then create mock observations by placing each galaxy at an identical distance to its observational counterpart (see Section~\ref{subsec:ss} for a sample matching strategy), observing them from the same directions as the optical images. 
The \HI integrated spectra resembling those from the ALFALFA survey are obtained by convolving them with a beam size of $212$ arcsec (approximating the Arecibo beam of $\sim3.8' \times 3.3'$) and a channel width of $5.5$ km/s, both consistent with the ALFALFA observational setup~\citep{2011AJ....142..170H,2018ApJ...861...49H}. We also apply a Hanning smooth, a standard spectral smoothing that reduces channel-to-channel noise fluctuations, yielding a final velocity resolution of $10$ km/s, and then add random Gaussian noise equivalent to the ALFALFA noise level. This \HI integrated spectra obtained allowed us to measure the \HI line width $V_{85}$.

An observation-equivalent definition of $M_\mathrm{HI}$ is required for the TNG100 galaxies. Although IllustrisTNG reproduces the overall \HI size–mass relation consistent with observations~\citep{2019MNRAS.490...96S}, using the total \HI mass together with the `true' \HI radius at which the \HI column density drops to $1,M_\odot,\mathrm{pc}^{-2}$ leads to a systematic offset from the empirical $M_\mathrm{HI}$–$R_\mathrm{HI}$ relation, because the total \HI mass includes a significant amount of diffuse halo gas. We therefore define the \HI mass as that enclosed within $2.2$ times the `true' \HI radius. With this definition, the simulated galaxies reproduce the observed $M_\mathrm{HI}$–$R_\mathrm{HI}$ relation of~\citet{2016MNRAS.460.2143W} with small scatter (see Appendix~\ref{appendix: calibration} for details).

We finally infer the observed-style \rhi from $M_\mathrm{HI}$ using their empirical correlation. We use these mock images and integrated spectra \HI to determine the dynamical mass and baryonic within \rhi using the exact same method as for observation, as described in Section~\ref{sec:method}. The baryonic mass is computed using the actual stellar mass within $5R_\mathrm{e}$ and the \HI mass within \rhi. This approach yielded an observed-style enclosed dark matter mass within \rhi for each galaxy in the sample. 

We define the values of the TNG100 mass properties on the basis of the particle counts as the ground truth for method validation. The true dynamical mass within \rhi is obtained by adding the masses of all particles (including stellar, gas and dark matter) within \rhi. We also identify the counterparts of the TNG100 galaxies in the dark-matter-only simulation `TNG100-1-dark' using the Subhalo Matching To Dark catalogue provided by the IllustrisTNG project. We compute the corresponding enclosed dark matter mass within \rhi and scale it to account for the difference in dark matter particle masses between the TNG100-dark and TNG100 simulations. This yields $M_\mathrm{dark}(<R_\mathrm{HI})$, representing the dark matter distribution independent of baryonic influence~\citep[e.g.][]{1996ApJ...462..563N,2008MNRAS.391.1685S}.

\subsection{Sample matching}
\label{subsec:ss}
The enclosed dark matter mass within \rhi is influenced not only by the dark matter halo mass, which correlates with stellar mass, but also by the enclosed radius \rhi itself. We therefore perform a nearest-neighbour one-to-one matching on the $M_* -R_\mathrm{HI}$ plane for the ALFALFA parent sample and the TNG100 \HI galaxies, requiring that the matched pairs differ by no more than 0.05 dex in $M_*$ and 0.02 dex in \rhi, to ensure a fair comparison between observational and simulated galaxies.

\begin{figure}
    \centering
    \includegraphics[width=0.9\hsize]{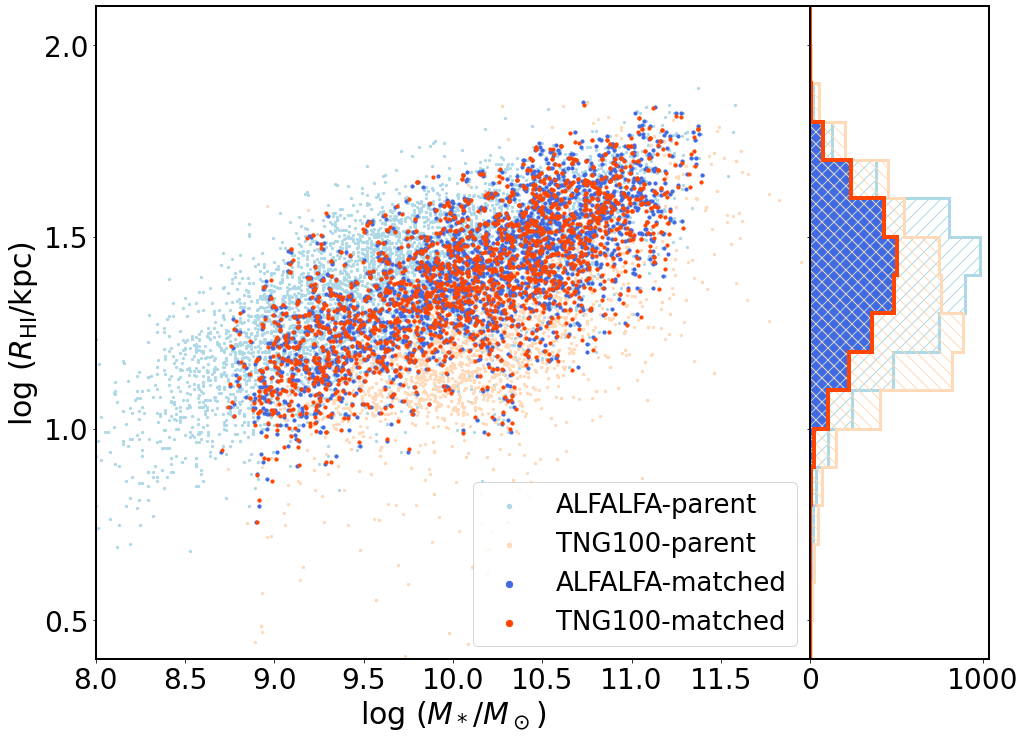}
    \caption{Left: the ALFALFA matched sample galaxies (blue) and the TNG100 matched sample (red) on the $\log(M_*) -\log(R_\mathrm{HI})$ plane. We plot the ALFALFA parent sample (light blue) and TNG100 \HI galaxies (orange) in the background. Right: $R_\mathrm{HI}$ distribution for the aforementioned samples plotted in the same colour scheme.}
    \label{fig:subsample}
\end{figure}

\begin{figure}
    \centering
    \includegraphics[width=0.9\hsize]{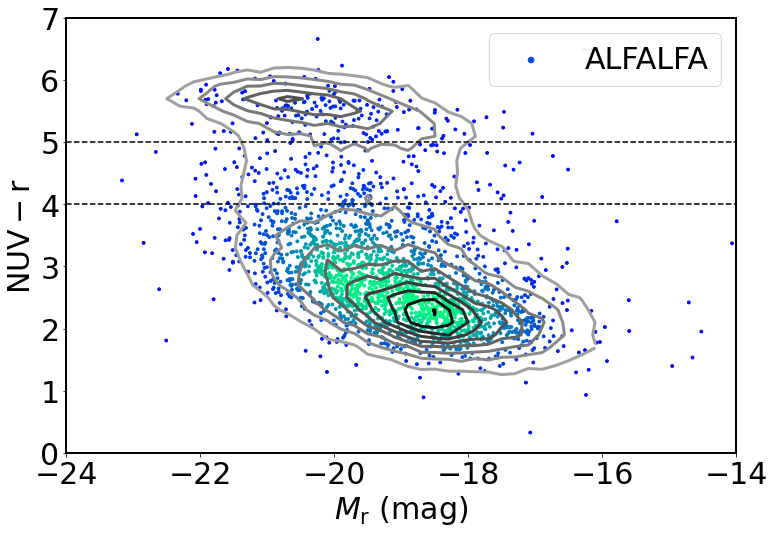}
    \caption{Colour-magnitude diagram for the sample galaxies. The ALFALFA sample is displayed with varying colours representing density and consists primarily of star-forming galaxies. Additionally, the NASA-Sloan Atlas catalogue is shown in the background with the contour lines~\citep{2011AJ....142...31B}.}
    \label{fig:cmd}
\end{figure}

\begin{figure}
    \centering
    \includegraphics[width=0.9\hsize]{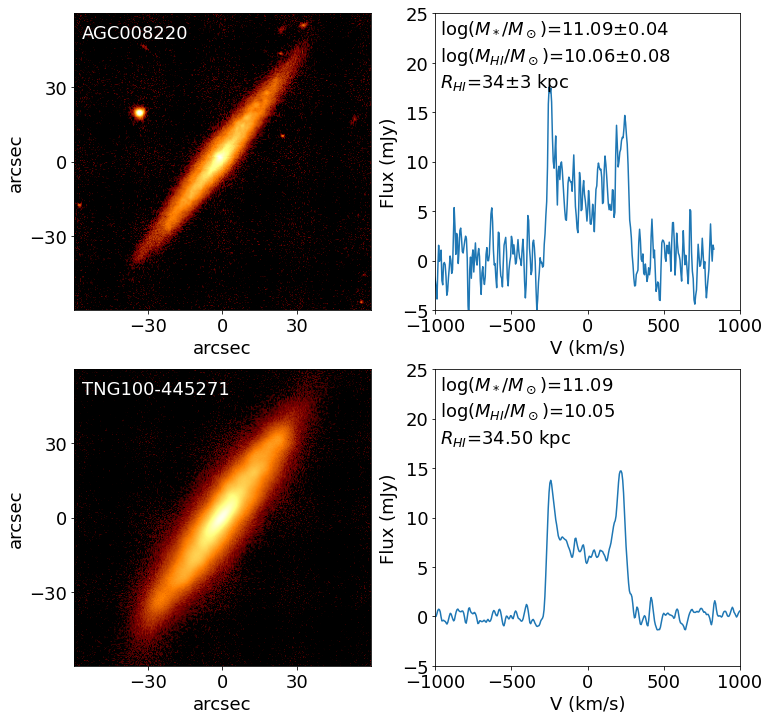}
    \caption{$r$-band images and \HI integrated spectra for an ALFALFA galaxy AGC008220 (top) and a mock galaxy TNG100-445271 (bottom) from the TNG100 simulation. The \HI spectra panels (right) display the corresponding stellar mass, \HI mass, and \rhi values for each galaxy.} 
    \label{fig:example}
\end{figure}

This matching results in a final sample comprising 2453 galaxies from the ALFALFA survey, each paired with a corresponding TNG100 galaxy. Figure~\ref{fig:subsample} demonstrates the stellar mass - \rhi distribution, along with histograms showing the \rhi distribution of this matching.
The matched ALFALFA sample has a stellar mass range from $10^{8.7}$ to $ 10^{11.4} M_\odot$, narrower than that of the ALFALFA parent sample. Figure~\ref{fig:cmd} shows the colour–magnitude diagram of the matched ALFALFA galaxies, indicating that the sample is dominated by star-forming galaxies.
Figure~\ref{fig:example} presents the $r$-band images and \HI integrated spectra for an ALFALFA galaxy and its corresponding mock galaxy from TNG100. The paired galaxies exhibit similar morphology and spectral features, and have similar stellar and \HI masses, resulting in comparable \HI radii.

\section{Results}
\label{sec:results}
In this section, we show the characteristics of the dynamical and dark matter mass within \rhi for the ALFALFA sample and the TNG100 sample with well-matched $M_*$ and \rhi. We divide both samples into five mass bins according to the stellar mass of the TNG100 sample in all matched pairs, ensuring an equal number of ALFALFA and TNG100 galaxies in each mass bin. The first four bins contain roughly 490--750 galaxies each, while the highest-mass bin is smaller, containing only 132 galaxies. In the following context, the results obtained from the mock observations of the TNG100 sample are denoted as the TNG100-mock, and the true values of the TNG100 sample are denoted as the TNG100-true.

\subsection{Dynamical mass}
\begin{figure*}
    \centering
    \includegraphics[width=\textwidth]{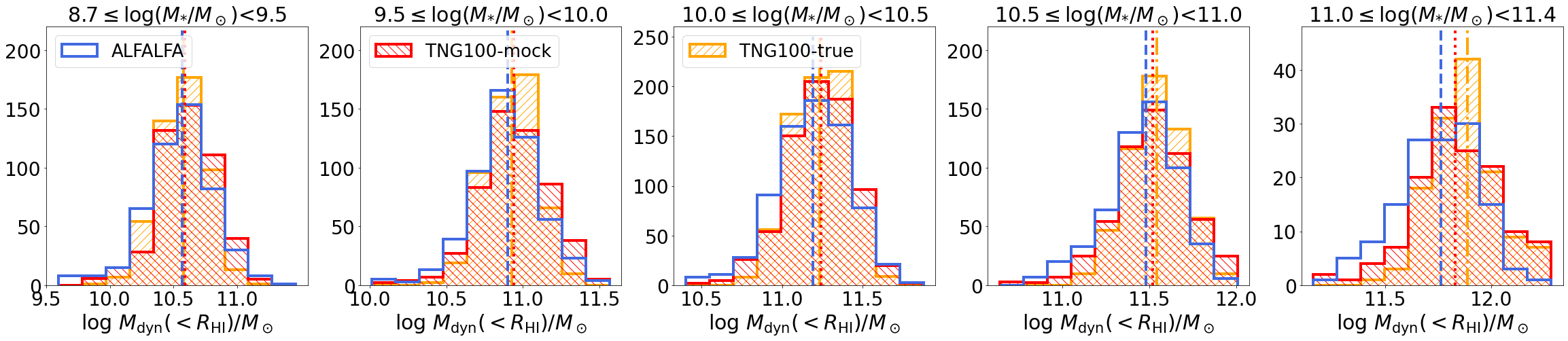}
    \caption{Histograms of the enclosed dynamical mass within \rhi for the ALFALFA sample and the TNG100 sample in five mass bins. The measurements for the ALFALFA sample are shown in blue, the true values of the TNG100 sample (TNG100-true) are plotted in orange, and the results obtained from the mock observations of the TNG100 sample (TNG100-mock) are shown in red. Median values for each dataset are indicated by a vertical dashed line matching the colour of the sample.} 
    \label{fig:dyn-hist}
\end{figure*}

We plot the histograms of the enclosed dynamical mass within \rhi for the ALFALFA sample and the TNG100 sample and show the median value for each data set in Figure~\ref{fig:dyn-hist}. We also perform pairwise statistical tests to assess differences between the distributions of the enclosed dynamical masses within \rhi for TNG100-mock and TNG100-true, and for ALFALFA and TNG100-mock across stellar mass bins, with the bin number indicated as well as the corresponding statistical results in Table~\ref{tab:com-dyn}. Specifically, we compute the Mann–Whitney–Wilcoxon\citep[MWU;][]{wilcoxon1945individual,mann1947test} p-value, which tests for a shift in the median between two samples; the Cliff's $\delta$~\citep{cliff1993dominance,cliff1996answering}, which quantifies the direction and magnitude of the distributional shift between the two samples; and the Kolmogorov–Smirnov~\citep[KS;][]{an1933sulla,smirnov1939estimate} p-value, which tests whether two samples are drawn from the same parent distribution. For the MWU and KS tests, a low p-value (typically $< 0.05$) indicates that the null hypothesis (the two samples are drawn from the same parent distribution) can be rejected with $\geq95\%$ confidence. In this work, we adopt a more conservative reference threshold of $0.01$ to highlight only the most robust differences. For Cliff's $\delta$, we follow standard interpretations, where $|\delta|<0.11$ corresponds to a negligible shift between distributions. These conventions allow us to distinguish statistically significant differences from negligible ones that are too small in magnitude to affect our scientific conclusions.
\begin{table*}
    \caption{Pairwise statistical tests of the enclosed dynamical masses within \rhi across stellar mass bins.}    
    \centering
    \begin{tabular}{ccccccc}
    \toprule
    \multirow{3}{*}{Comparison} & \multirow{3}{*}{Statistics} & \multicolumn{5}{c}{$\log(M_*/M_\odot)$} \\
    \cline{3-7}
    & & $[8.7,9.5)$ & $[9.5,10.0)$ & $[10.0,10.5)$ & $[10.5,11.0)$ & $[11.0,11.4)$ \\
    \cline{3-7}
    & & $N=491$ & $N=532$ & $N=747$ & $N=551$ & $N=132$ \\
    \midrule
    \multirow{3}{*}{TNG100-mock vs TNG100-true}  & MWU p-value & 0.013 & 0.088 & 0.763 & 0.093 & 0.137 \\
     & Cliff's $\delta$ & 0.091 & 0.060 & 0.009 & -0.058 & -0.106 \\
     & KS p-value & 6.1e-04 & 7.0e-04 & 0.235 & 0.036 & 0.130 \\
    \cline{1-7}
    \multirow{3}{*}{ALFALFA vs TNG100-mock} & MWU p-value & 0.006 & 3.1e-04 & 2.4e-04 & 6.6e-05 & 7.0e-04 \\
     & Cliff's $\delta$ & -0.102 & -0.128 & -0.110 & -0.139 & -0.241 \\
     & KS p-value & 0.015 & 3.3e-04 & 0.003 & 9.3e-04 & 0.017 \\
    \cline{1-7}
    \bottomrule
    \end{tabular}
    \label{tab:com-dyn}
    \tablefoot{The samples compared are: TNG100-mock versus TNG100-true, and ALFALFA versus TNG100-mock. The number of galaxies in each bin is indicated in the third row of the table. For each comparison, we report the MWU p-value, and Cliff’s $\delta$, and KS p-value. In both MWU and KS tests, small p-values (e.g., $p<0.01$, $\sim 2.6\sigma$ significance) indicate that the null hypothesis can be rejected, i.e. that the two samples are unlikely to share the same median (MWU) or to be drawn from the same distribution (KS). Cliff’s $\delta$ quantifies the direction and magnitude of the distributional shift: positive values indicate that the first sample tends to be larger, and vice versa; $|\delta|\gtrsim0.11,\,0.28,\,0.43$ correspond to small, medium, and large effects.}
\end{table*}

The method for measuring dynamical mass introduces minimal bias: compared to TNG100-true, the median dynamical masses within \rhi for TNG100-mock are higher by $<3\%$ in the first three stellar-mass bins, lower by $5.3\%$ in the fourth bin, and lower by $12.5\%$ in the highest-mass bin. This conclusion is supported by statistical tests: the MWU p-values exceed 0.01 and the absolute values of Cliff’s $\delta$ are below 0.11 in all bins, indicating negligible systematic shifts in the median values. Additionally, the KS p-values in the three massive bins are higher than 0.01, providing further evidence that the distributions of dynamical masses within \rhi for TNG100-mock and TNG100-true are statistically consistent in these bins. For the two lowest-mass bins, the KS p-values are lower than 0.01, indicating that the full distributions are not identical; however, the median values differ by less than 3\%, suggesting that these distributional discrepancies do not introduce a significant systematic bias in the measurement of dynamical mass.

The median dynamical masses within \rhi for the ALFALFA sample are systematically lower than those of the TNG100-mock across five stellar mass bins, with the difference minimal in the lowest mass bin and increasing towards higher mass bins. Pairwise MWU tests and Cliff's $\delta$ indicate only slight shifts in low-mass bins (small effect, $|\delta|\lesssim0.11$; MWU $p<0.01$), with negative $\delta$ values showing that ALFALFA tends to have lower dynamical masses. In the highest mass bin the shift becomes more pronounced ($\delta=-0.24$), largely driven by $\sim$21\% of ALFALFA galaxies having $M_\mathrm{dyn}(<R_\mathrm{HI})<10^{11.6} M_\odot$ compared to 9\% of TNG100-mock galaxies with such low dynamical masses. KS tests further confirm the trend of increasing differences with stellar mass: in the lowest-mass bin, the difference is relatively small ($p\sim0.11$), while in the intermediate bins the differences are statistically significant ($p<0.01$). The higher p-value in the highest-mass bin ($p\sim0.17$) is mainly due to the small sample size in this bin (132 galaxies, less than 25\% of the other bins) and does not contradict the overall trend of larger differences at higher stellar mass.

\subsection{Dark matter halo mass}
\begin{figure*}
    \centering
    \includegraphics[width=\textwidth]{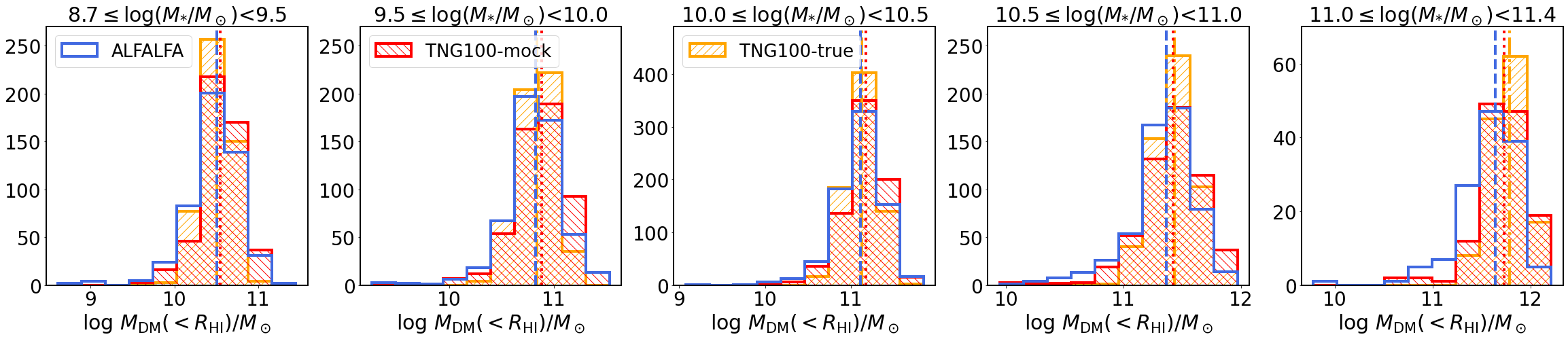}
    \caption{Histograms of the enclosed dark matter mass within \rhi for the ALFALFA sample and the TNG100 sample in five mass bins. The measurements for the ALFALFA sample are shown in blue, the true values of the TNG100 sample (TNG100-true) are plotted in orange, and the results obtained from the mock observations of TNG100 sample (TNG100-mock) are shown in red. Median values for each dataset are indicated by a vertical dashed line matching the colour of the sample.}
    \label{fig:dm-hist}
\end{figure*}

We show the histograms of the enclosed dark matter mass within \rhi for the ALFALFA sample and the TNG100 sample in Figure~\ref{fig:dm-hist}. We also report MWU p-values, Cliff's $\delta$, and KS p-values for the distributions of the enclosed dark matter masses within \rhi, comparing TNG100-mock with TNG100-true and ALFALFA with TNG100-mock across stellar mass bins, with the bin number and the corresponding statistical results shown in Table~\ref{tab:com-dm}.
\begin{table*}
    \caption{Pairwise statistical tests of the enclosed dark matter masses within \rhi across stellar mass bins. }
    \centering
    \begin{tabular}{ccccccc}
    \toprule
    \multirow{3}{*}{Comparison} & \multirow{3}{*}{Statistics} & \multicolumn{5}{c}{$\log(M_*/M_\odot)$} \\
    \cline{3-7}
    & & $[8.7,9.5)$ & $[9.5,10.0)$ & $[10.0,10.5)$ & $[10.5,11.0)$ & $[11.0,11.4)$ \\
    \cline{3-7}
    & & $N=491$ & $N=532$ & $N=747$ & $N=551$ & $N=132$ \\
    \midrule
    \multirow{3}{*}{TNG100-mock vs TNG100-true} & MWU p-value & 9.2e-07 & 3.2e-06 & 2.1e-04 & 0.774 & 0.193 \\
     & Cliff's $\delta$ & 0.181 & 0.165 & 0.111 & -0.010 & -0.093 \\
     & KS p-value & 1.3e-07 & 2.6e-09 & 7.8e-05 & 0.060 & 0.036 \\
    \cline{1-7}
    \multirow{3}{*}{ALFALFA vs TNG100-mock} & MWU p-value & 2.3e-04 & 1.3e-05 & 1.0e-05 & 1.2e-06 & 8.8e-05 \\
     & Cliff's $\delta$ & -0.136 & -0.155 & -0.132 & -0.169 & -0.279 \\
     & KS p-value & 0.002 & 8.7e-05 & 1.2e-04 & 5.5e-05 & 0.002 \\
    \cline{1-7}
    \bottomrule
    \end{tabular}
    \label{tab:com-dm}
    \tablefoot{The samples compared are: TNG100-mock versus TNG100-true, and ALFALFA versus TNG100-mock. The number of galaxies in each bin is indicated in the third row of the table. Test statistics are the same as in Table~\ref{tab:com-dyn}.}
\end{table*}

The enclosed dark matter mass within \rhi for the TNG100-mock is generally consistent with that of the TNG100-true in the two high-mass bins, suggesting that methodological biases in the dark matter measurements are small for massive galaxies. In these high-mass bins, MWU p-values are higher than $0.01$ with $|\delta|<0.11$, and KS p-values exceed $0.01$, indicating negligible systematic shifts in the median values and statistically consistent distributions. In contrast, in the three low-mass bins ($M_*\lesssim10^{10.5}M_\odot$), the TNG100-mock tends to yield slightly higher enclosed dark matter masses than the true values. This is reflected by MWU p-values well below 0.01 with positive $\delta$ ($0.11$--$0.18$), and very small KS p-values ($p<10^{-4}$), indicating statistically significant shifts in the full distributions. This discrepancies reflect differences in measurement methodology and mass definitions in the low-mass bins.

Mirroring the trend observed in the dynamical masses within \rhi, the median enclosed dark matter masses within \rhi for the ALFALFA sample are systematically lower than those of the TNG100-mock across all five stellar mass bins, with the difference minimal in the lowest mass bin and increasing toward higher bins. This is supported by MWU p-values $<0.01$ in every mass bin, while Cliff's $\delta$ is consistently negative, with magnitudes increasing from $0.13$ to $0.28$ toward higher stellar mass. This increasing offset toward higher stellar masses in the median is caused by a tail of galaxies with comparatively low dark matter content, which becomes more prominent at higher stellar masses. This effect is particularly evident in the highest mass bin ($M_*>10^{11}M_\odot$), where $\sim$32\% of ALFALFA galaxies have $M_\mathrm{DM}(<R_\mathrm{HI})<10^{11.5}M_\odot$, compared to only 17\% of the TNG100-mock galaxies. KS test p-values are below 0.01 in all mass bins, confirming that the ALFALFA and TNG100-mock samples follow statistically distinct distributions. We note that although the comparison between TNG100-mock and TNG100-true shows that the low-mass bins are affected by methodological biases, the comparison between ALFALFA and TNG100-mock in this regime is still informative because both datasets are processed with the same observational method. The interpretation in this mass range, however, must account for the fact that the differences cannot be attributed purely to physical origins.

\begin{figure}
    \centering
    \includegraphics[width=0.9\hsize]{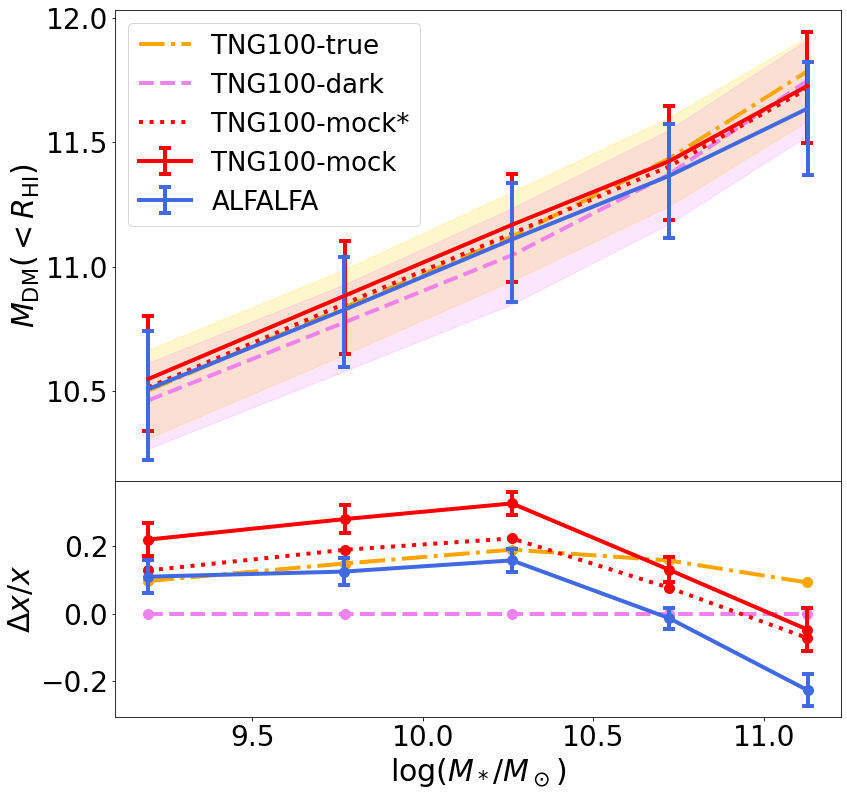}
    \caption{Relation between the stellar mass and the enclosed dark matter mass within \rhi. The upper panel shows the median and the $1\sigma$ scatter (16th to 84th percentile) for the ALFALFA sample galaxies (blue solid), the mock observations (TNG100-mock, red solid) and the true value (TNG100-true, orange dash-dotted) of the TNG100 sample, the counterpart of the sample from the TNG100-dark simulation (pink dashed), and a `dark matter' mass derived by removing the true baryonic mass of the TNG100 from the mock dynamical mass of the TNG100-mock (TNG100-mock*, red dotted). The lower panel presents the relative change of the median values compared to that of the TNG100-dark line in the same colour scheme and line style. We also indicate the errors for the median values of the ALFALFA sample and TNG100-mock with their corresponding error bars.}
    \label{fig:dm-relation}
\end{figure}

We further quantify the difference between these histograms and present them as a relation between the stellar mass and the dark matter halo mass within \rhi, as shown in Figure~\ref{fig:dm-relation}. The upper panel shows the median and $1\sigma$ scatter (16th to 84th percentile) for each mass bin. Alongside the ALFALFA and the TNG100 sample, we include the enclosed dark matter mass within \rhi for the counterpart in the TNG100-dark simulation (denoted as TNG100-dark). We also include a `dark matter' mass within \rhi that is derived by removing the true baryonic mass of the TNG100-true from the mock dynamical mass of the TNG100-mock (denoted as TNG100-mock*). The lower panel shows the relative change of the median values compared to that of the TNG100-dark for each mass bin in the same colour scheme and line style. 

Across five mass bins, the dark matter masses within \rhi of the TNG100-true are about 10-20\% higher than those of the TNG100-dark. This suggests an overall halo contraction phenomenon in the TNG100 simulation, as reported in~\citet{2018MNRAS.481.1950L}.

The TNG100-mock exhibits higher enclosed dark matter masses within \rhi than the TNG100-true in the three low-mass bins, while being generally consistent with the TNG100-true in the two high-mass bins, as shown by the statistical results in the previous section. In the highest mass bin ($M_* \geq 10^{11} M_\odot$), the enclosed dark matter mass within \rhi of the TNG100-mock is slightly lower ($\sim$13\%) than that of the TNG100-true.  
To investigate the origin of these discrepancies, we introduce the TNG100-mock*, defined by subtracting the true baryonic mass of the TNG100-true (summing all stellar and gas particles) from the measured dynamical mass of the TNG100-mock obtained using the observational method.  
In the low-mass bins, the TNG100-mock* closely matches the TNG100-true values and lies below the TNG100-mock, indicating that the discrepancies in these bins primarily arise from methodological biases in baryonic mass estimation. Our observationally motivated approach (Section~\ref{sec:method}, Eq.~\ref{equ:bary}) includes stellar mass and cold gas phases (atomic and molecular gas) but omits hot gas. However, TNG100 galaxies, especially low-mass, gas-rich ones, contain a non-negligible fraction of hot gas, which is observationally inaccessible and thus not accounted for in our gas mass estimates~\citep{2018MNRAS.477..450N,2020MNRAS.499..768Z}.
In the high-mass bins, the overall gas fraction decreases with increasing stellar mass, reducing the difference between TNG100-mock* and TNG100-mock. Meanwhile, a modest discrepancy between TNG100-mock* and TNG100-true persists and becomes slightly more pronounced toward higher stellar mass. This arises from a slight underestimation of the dynamical mass within \rhi, which is calculated from the \HI line width $V_{85}$ and the disc inclination $i$. Factors such as misalignment between the \HI and stellar discs, underestimation of the intrinsic flattening $q_0$ of the stellar disc, or outliers can lead to a slightly lower circular velocity $V_\mathrm{c}$, and hence $M_\mathrm{dyn}(<R_\mathrm{HI})$. While this effect exists at all masses, the limited sample size in the highest-mass bin (132 galaxies, less than 25\% of other bins) makes the average results more sensitive to individual measurement uncertainties.

The ALFALFA sample exhibits systematically lower enclosed dark matter masses within \rhi than the TNG100-mock, but the physical interpretation differs across the mass range. In the low-mass bins, observational measurements may suffer from unaccounted hot gas components, leaving open whether the enclosed dark matter masses in ALFALFA are closer to those in TNG100-true (or TNG100-mock*) or even as low as in TNG100-dark. At higher stellar masses, the gas fraction decreases such that the differences are no longer mainly driven by potential biases from the unaccounted hot gas. In the highest mass bin ($M_* \geq 10^{11}M_\odot$), the total gas-to-dark matter ratio within \rhi in TNG100 is only $\sim0.05$ (similar to the cold gas fraction in ALFALFA), indicating that gas has a negligible impact on the enclosed mass budget. In this bin, the enclosed dark matter mass within \rhi in ALFALFA is about 19\% lower than in the TNG100-mock and 23\% lower than in the TNG100-dark. Since both ALFALFA and TNG100-mock are subject to the same dynamical mass estimation bias, this discrepancy reflects a robustly lower enclosed dark matter mass in ALFALFA compared to TNG100, which would only become more pronounced if additional unaccounted gas were present in the ALFALFA.

We find that in the highest mass bin in the ALFALFA sample, all of the 131 galaxies with stellar mass exceeding $10^{11}M_\odot$ are classified as late-type based on our visual inspection of their images, with $R_\mathrm{HI}$ (i.e. the radii enclosing the dark matter traced by \HI) ranging from 20 to 70 kpc. The difference we found between ALFALFA and TNG100 galaxies at this mass range can hardly be explained solely by baryonic feedback processes. Previous studies have shown that supernova-driven feedback can form dark matter cores of a few kpc in low-mass galaxies~\citep[e.g.][]{2012MNRAS.421.3464P,2020MNRAS.491.4523F}, and AGN-driven outflows are able to generate cores up to about 10 kpc in intermediate-mass galaxies~\citep[e.g.][]{2013MNRAS.432.1947M,2023MNRAS.518.5356L}. However, forming cores as large as 20–70 kpc in galaxies with $M_* > 10^{11} M_\odot$ remains challenging, as it would require substantially more efficient energy transfer to the dark matter halo than currently demonstrated. At this high-mass end, the dark matter mass within \rhi is an indicator of the total dark matter halo mass, thus the lack of galaxies with low enclosed dark matter mass within \rhi in the TNG100 simulation highlights a deficiency of massive galaxies residing in small dark matter haloes in the TNG100 simulation: In the ALFALFA sample, 32\% of massive late-type galaxies reside in smaller dark matter halos (below $10^{11.5} M_\odot$ within \rhi), whereas in the TNG100 simulation, this figure is 17\%.

\section{Discussion}
\label{sec:discussion}
In our study, we find that although a larger fraction of massive late-type galaxies ($10^{11}M_\odot \leq M_* < 10^{11.4}M_\odot$) reside in relatively smaller dark matter haloes compared to the TNG100 simulation, the majority still have halo masses consistent with those predicted by TNG100. This implies that, overall, the SHMR of these galaxies agrees reasonably well with the TNG100 relation. At first sight, this appears to be at odds with the compilation of \citet{2025ApJ...980..233W}, which compared the SHMRs of late-type and early-type galaxies across a range of observations and simulations. In their sample, more than 95\% of the 22 late-type galaxies in this stellar mass range-measured from \HI rotation curves~\citep{2019A&A...626A..56P,2023MNRAS.518.6340D} fall below the $1\sigma$ scatter of the IllustrisTNG-predicted SHMR, which does not differentiate between morphological types.

\begin{figure}
    \centering
    \includegraphics[width=0.9\hsize]{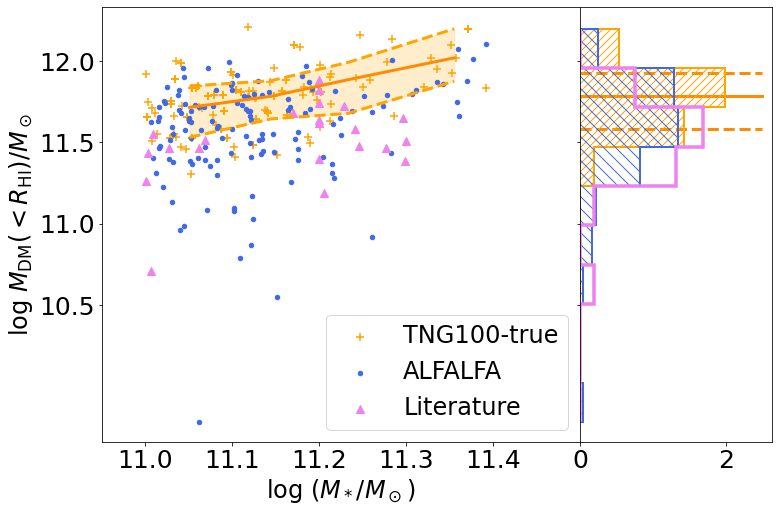}
    \caption{Left: relation between the stellar mass and the enclosed dark matter mass within \rhi within the stellar mass range $10^{11}M_\odot \leq M_*<10^{11.4} M_\odot$. We show the ALFALFA sample in blue, the TNG100-true in orange, and in pink the combined literature late-type galaxies from~\citet{2019A&A...626A..56P} and~\citet{2023MNRAS.518.6340D}. The orange solid and dashed lines indicates the median and the $1\sigma$ scatter (16th to 84th percentile) for TNG100-true. Right: normalized distributions of the enclosed dark matter mass within $R_\mathrm{HI}$ for the same samples, plotted in the same colour scheme.}
    \label{fig:others-rhi}
\end{figure}

To clarify the observed discrepancy, we compute the enclosed mass within \rhi for 15 late-type galaxies from~\citet{2019A&A...626A..56P} and 7 late-type galaxies from~\citet{2023MNRAS.518.6340D} within this stellar mass range. These are plotted alongside the measurements of the ALFALFA sample and the true values of the TNG100 sample, as illustrated in Figure~\ref{fig:others-rhi}. We find that only 73\% of the 22 galaxies from~\citet{2019A&A...626A..56P} and~\citet{2023MNRAS.518.6340D} fall below the $1\sigma$ scatter of the TNG100-true, indicating a systematic offset less pronounced than the $1\sigma$ discrepancy reported by~\citet{2025ApJ...980..233W}. Nevertheless, the distribution of these galaxies in the literature is covered by the ALFALFA sample, which largely overlaps with the TNG100-true.

Given the small size of the samples of~\citet{2019A&A...626A..56P} and~\citet{2023MNRAS.518.6340D}, sampling fluctuations can partly account for the apparent offset. We perform a bootstrap test by randomly drawing 22 galaxies from the ALFALFA sample 10,000 times and computing the median of each draw. The median of the literature sample falls around the 4th percentile of the bootstrap distribution ($p\approx0.04$), consistent with being within $2\sigma$. This suggests that the lower dark matter masses reported by~\citet{2019A&A...626A..56P} and~\citet{2023MNRAS.518.6340D} could plausibly arise from sampling variance, although a mild intrinsic offset cannot be ruled out. In addition, systematic differences in the assumed halo density profiles may contribute to the lower observationally inferred $M_{200}$. Observational estimates typically extrapolate from the inner regions assuming NFW ($\gamma = 1$) or cored ($\gamma \sim 0$) profiles, which are shallower than the inner density slopes of simulated halos~\citep[$\gamma \sim 1.6$–$1.8$;][]{2018MNRAS.481.1950L}. A shallower profile implies a more slowly increasing cumulative mass with radius, leading to a lower extrapolated $M_{200}$ for a given enclosed mass within $R_\mathrm{HI}$. Therefore, part of the offset between the literature galaxies and the TNG100 relation may reflect this intrinsic difference in halo structure, in addition to sample variance.

\begin{figure}
    \centering
    \includegraphics[width=0.9\hsize]{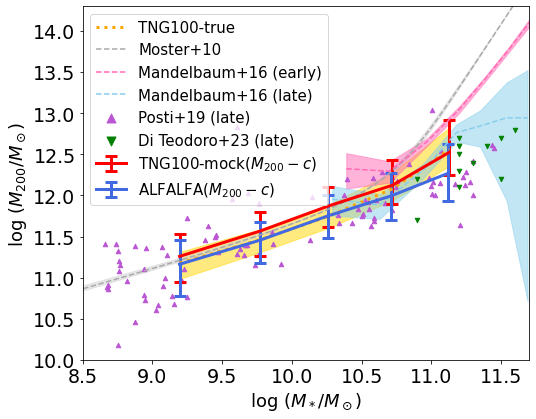}
    \caption{Relation between the stellar mass and the extrapolated dark matter halo mass. We show the median and the $1\sigma$ scatter (16th to 84th percentile) for ALFALFA sample galaxies (blue solid), TNG100-mock (red solid), and TNG100-true (orange dashed and shadow). The estimated dark matter halo mass $M_\mathrm{200}$ is derived from the dark matter mass within \rhi, using an NFW profile and the halo mass-concentration relation from~\citet{2014MNRAS.441.3359D}. Late-type galaxies from ~\citet{2019A&A...626A..56P} and ~\cite{2023MNRAS.518.6340D} are shown with purple and green triangles, respectively. The SHMRs for early-type (pink dashed and shadow) and late-type (light blue dashed and shadow) galaxies from~\citet{2016MNRAS.457.3200M} are also displayed.}
    \label{fig:ext-m200}
\end{figure}

We estimate the dark matter halo mass $M_\mathrm{200}$ by extrapolating the dark matter mass within \rhi using an NFW profile~\citep{1996ApJ...462..563N,1997ApJ...490..493N}, along with the halo mass-concentration relation proposed in~\citet{2014MNRAS.441.3359D}, applicable to both the ALFALFA sample and mock observations of TNG100 galaxies. $M_{200}$ is defined as the enclosed mass within the virial radius $r_{200}$, where the average density within $r_{200}$ is 200 times the critical density ($\rho_\mathrm{crit} = 1.37\times 10^{-7} \mathrm{M_\odot/pc^3}$, adopting a Hubble constant $H_0 = 70 \mathrm{km/s/Mpc}$); and the concentration is defined as the ratio of the virial radius $r_{200}$ and the scale radius of the NFW profile. Figure~\ref{fig:ext-m200} illustrates the relation between stellar mass and the extrapolated halo mass. We show the ALFALFA sample from this work as well as the true values and mock-based results of TNG100, with the SHMR for early-type galaxies and late-type galaxies as reported by~\citet{2016MNRAS.457.3200M} as a reference.
We note that the consistency between TN100-mock and TNG100-true confirms that the TNG100 galaxies follow the halo mass-concentration relation as described by~\citet{2014MNRAS.441.3359D}. By employing the same assumptions for estimating dark matter halo mass, the $M_*-M_\mathrm{200}$ relation for both ALFALFA and TNG100-mock galaxies reflects the $M_*-M_\mathrm{DM}(<R_\mathrm{HI})$ relation illustrated in Figure~\ref{fig:dm-relation}. In the mass range below $10^{11}M_\odot$, the ALFALFA and TNG100 samples are $1\sigma$ compatible with the SHMR for late-type galaxies found by~\citet{2016MNRAS.457.3200M}, whereas in the mass range above $10^{11}M_\odot$, the SHMR of late-type galaxies reported by~\citet{2016MNRAS.457.3200M} only aligns $1\sigma$ with the TNG100 galaxies and is slightly higher than the ALFALFA sample.

\section{Summary}
\label{sec:summary}
In this study, we evaluate the dark matter content in a sample of 2453 galaxies from the ALFALFA survey. We construct a one-to-one sample matched in the $M_* -R_\mathrm{HI}$ plane from the TNG100 simulation. We perform mock images and \HI observations under the conditions of their counterparts, enabling us to measure their dark matter content through the observational method. The main results are as follows:
   \begin{enumerate}
      \item Comparing the true values of the TNG100 sample with results derived from mock observations, we demonstrate that the method introduces minimal bias in the dynamical mass measurements, within about 5\% across most of the stellar mass range and up to 12.5\% at the highest masses ($M_* \geq 10^{11}M_\odot$). Statistical tests confirm that these offsets are not significant: the MWU p-values exceed 0.01 and the absolute values of Cliff’s $\delta$ are below 0.11 in all bins, indicating that the method provides reliable dynamical mass estimates.

      \item The median dark matter content of the ALFALFA sample is lower than that derived from the mock observations of the TNG100 simulation across all stellar mass bins. In each bin, this offset arises from a tail of galaxies with comparatively low dark matter content, and the prominence of this tail increases toward higher stellar masses. Statistical tests corroborate this trend: the MWU p-values are $<0.01$ in each mass bin, while Cliff’s $\delta$ is consistently negative, reflecting the systematically lower dark matter content in ALFALFA relative to the TNG100-mock. At the low-mass end ($M_* < 10^{9.5} M_\odot$), the difference is modest, with broadly consistent distributions between the two samples. At the high-mass end ($M_* > 10^{11} M_\odot$), the median dark matter content in ALFALFA is about 23\% lower than in the TNG100-dark simulation. Within this bin, about 32\% of ALFALFA galaxies have dark matter halo masses below $10^{11.5} M_\odot$ within \rhi, compared to only 17\% in the TNG100 mock galaxies. Considering that galaxies in the highest mass bin of the ALFALFA sample are late-type with the enclosed radii ranging from 20 kpc to 70 kpc, our findings suggest that a higher portion of these massive late-type galaxies reside in smaller dark matter haloes compared to TNG100 galaxies.
   \end{enumerate}

\section{Data availability}
Measurements for the ALFALFA parent sample are provided in the table \texttt{alfa4844.txt}, while the true values and mock results for the TNG100 sample are provided in \texttt{tng2453.txt}. Both tables are
only available in electronic form at the CDS via anonymous ftp to \url{cdsarc.u-strasbg.fr} (130.79.128.5) or via the CDS web interface at \url{http://cdsweb.u-strasbg.fr/cgi-bin/qcat?J/A+A/}.

\begin{acknowledgements}

      MY acknowledges the support of the National Natural Science Foundation of China, \emph{Young Scientists Fund} project, number 12303014.
      LZ acknowledges the support of the CAS Project for Young Scientists in Basic Research under grant No. YSBR-062 and National Key R\&D Program of China No. 2022YFF0503403.
      MY and LZ acknowledge the hospitality of the International Centre of Supernovae (ICESUN), Yunnan Key Laboratory at Yunnan Observatories Chinese Academy of Sciences.
      N.K.Y. is supported by the project funded by China Postdoctoral Science Foundation No. 2022M723175 and GZB20230766. 
      ZZ is supported by National Key R\&D Program of China No. 2023YFC2206403 and 2024YFA1611602. ZZ is also supported by NSFC grants No. 12588202, 12373012, 12041302 and CMS-CSST-2025-A08.
      
\end{acknowledgements}

\bibliographystyle{aa} 
\bibliography{ref.bib}

\begin{appendix}
\section{Method validation in TNG100}
\label{appendix: tng100}
In this section, we present the validation results of our observational method. We select galaxies from the TNG100 simulations and create mock \HI observations for them. The mock \HI integrated spectra are created using the procedure described in Section~\ref{sec:dands}. Each galaxy is positioned at a distance of 100 Mpc and is observed under conditions similar to those of the Green Bank Radio Telescope, which features a beam size of $540$ arcsec and a channel width of $1.4$ km/s. 

\begin{figure}
    \centering
    \includegraphics[width=\hsize]{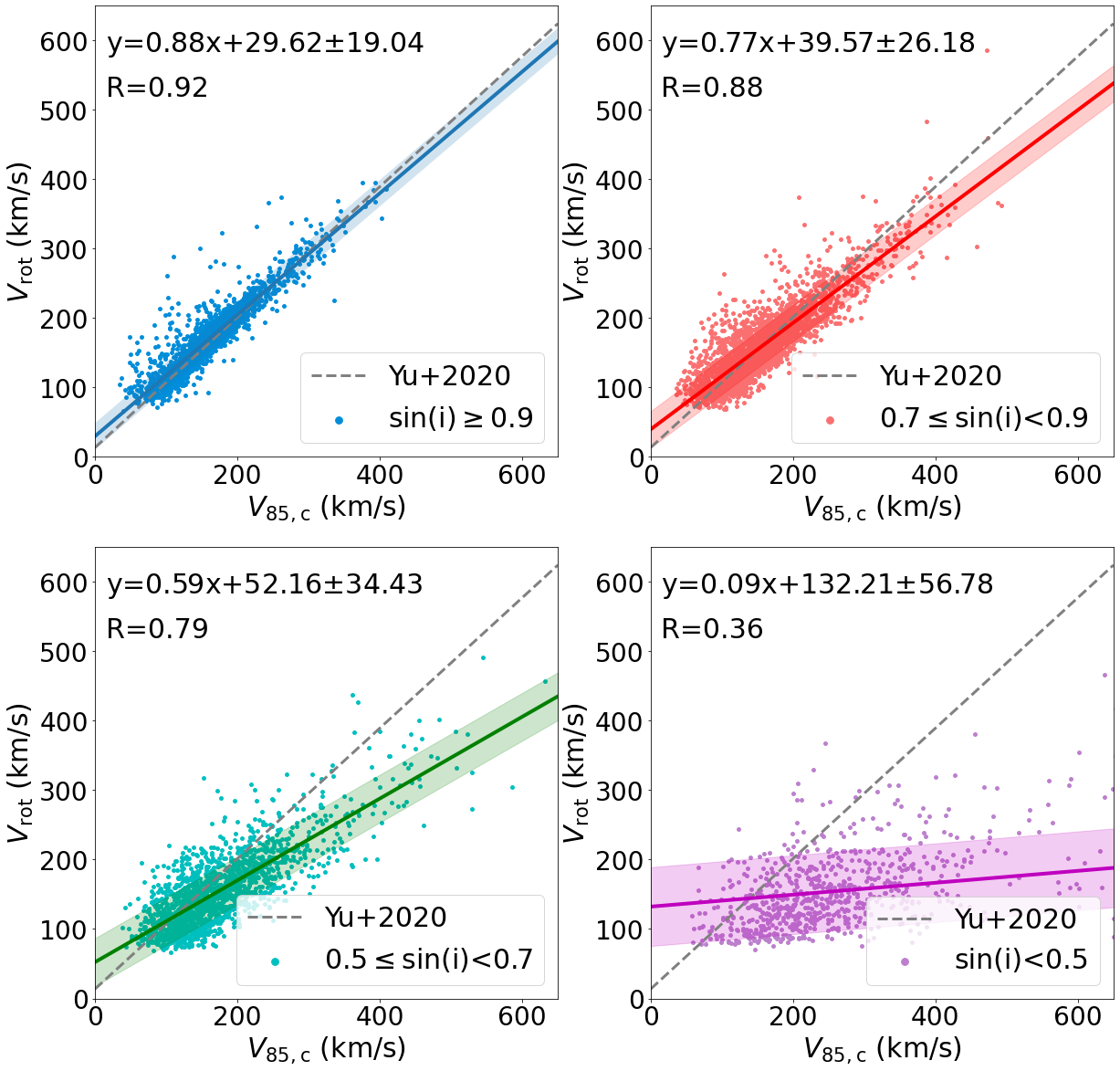}
    \caption{The relationship between rotational and corrected line width indicators for galaxies with different predicted inclinations. In each panel, the solid line and the corresponding shaded area represent the best-fitting line and its $1\sigma$ scatter, and the grey dashed line represents the relationship obtained by~\citet{2020ApJ...898..102Y}.
    }
    \label{fig:V85-Vrot}
\end{figure}

Following the method described in Section~\ref{sec:method}, we first measure the corrected line width indicator $V_\mathrm{85,c}$ and show its relationship to the true rotational velocity $V_\mathrm{rot}$ in Figure~\ref{fig:V85-Vrot}. The \HI galaxies are divided into four categories by their predicted inclination measured from the observational axis $q$. We fit a linear scaling relation between $V_\mathrm{85,c}$ and $V_\mathrm{rot}$ for each category and compare it with the scaling relation reported in~\citet{2022ApJS..261...21Y}. The findings indicate that only edge-on galaxies in the TNG100 adhere to this observed scaling relation.

\begin{figure}
    \centering
    \includegraphics[width=\hsize]{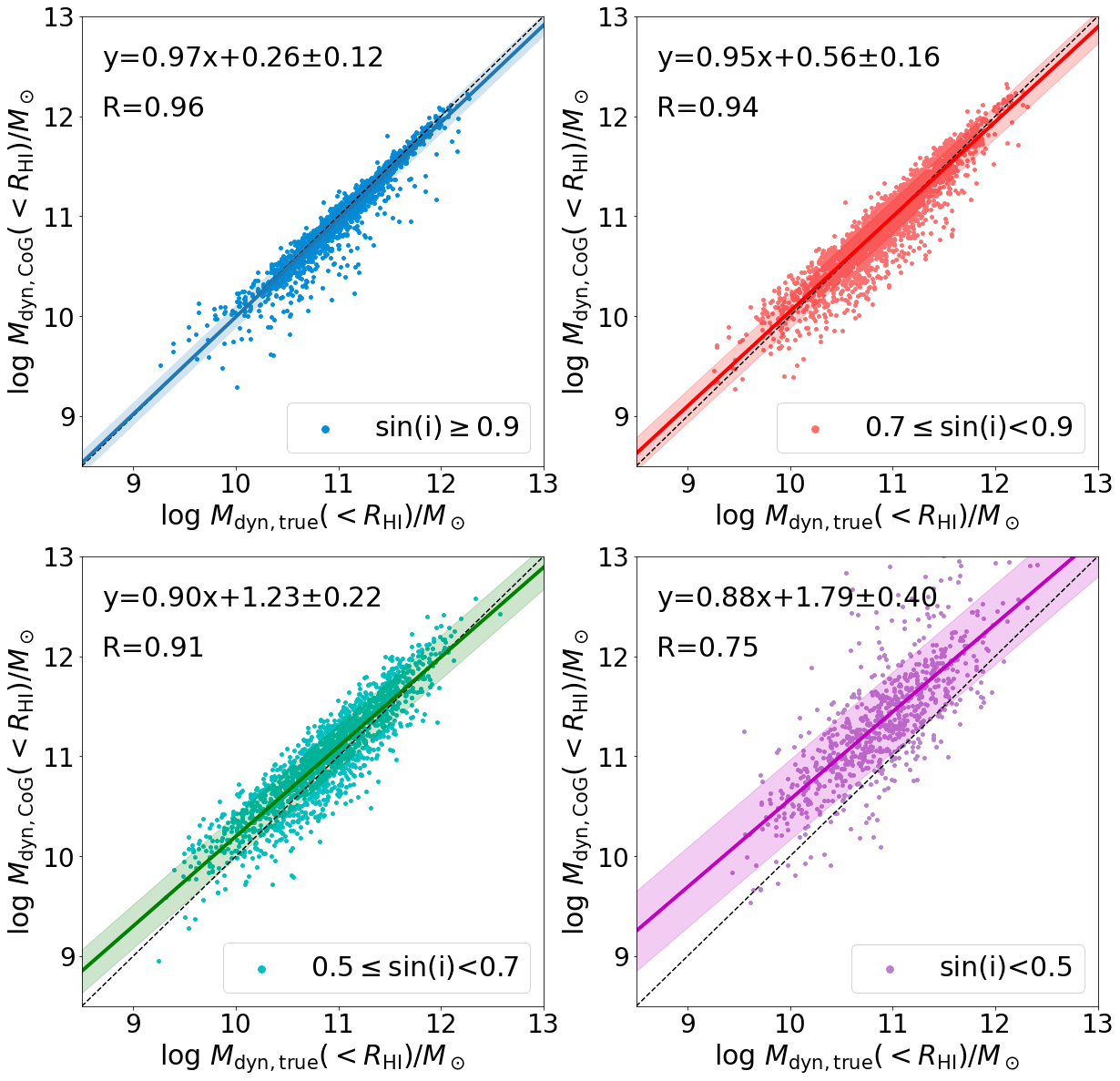}
    \caption{The relationship between the true total masses and measured dynamical masses from mock observations for the TNG100 parent sample. In each panel, the solid line and the corresponding shaded area represent the best-fitting line and its $1\sigma$ scatter, and the grey dashed line indicates the one-to-one line.}
    \label{fig:Mtrue-Mcog}
\end{figure}

We then calculate the circular velocity using the scaling relation reported in~\citet{2022ApJS..261...21Y} and from this we derive the dynamical mass. We show the comparison of this derived dynamical mass with the true dynamical mass in Figure~\ref{fig:Mtrue-Mcog}. We establish a linear scaling relationship for the four categories of different inclinations. The findings indicate that this method can accurately recover the dynamical mass of galaxies without significant bias within the specified stellar mass range, provided that these galaxies have an inclination of $\sin(i)\geq 0.7$ ($i \gtrsim 45^\circ$).

Therefore, we finally select galaxies with $\sin(i)\geq 0.7$ to form the parent sample of TNG100.

\section{Spectra of ALFALFA parent sample}
\label{appendix: alfa_sample}
\begin{figure*}
    \centering
    \includegraphics[width=\textwidth]{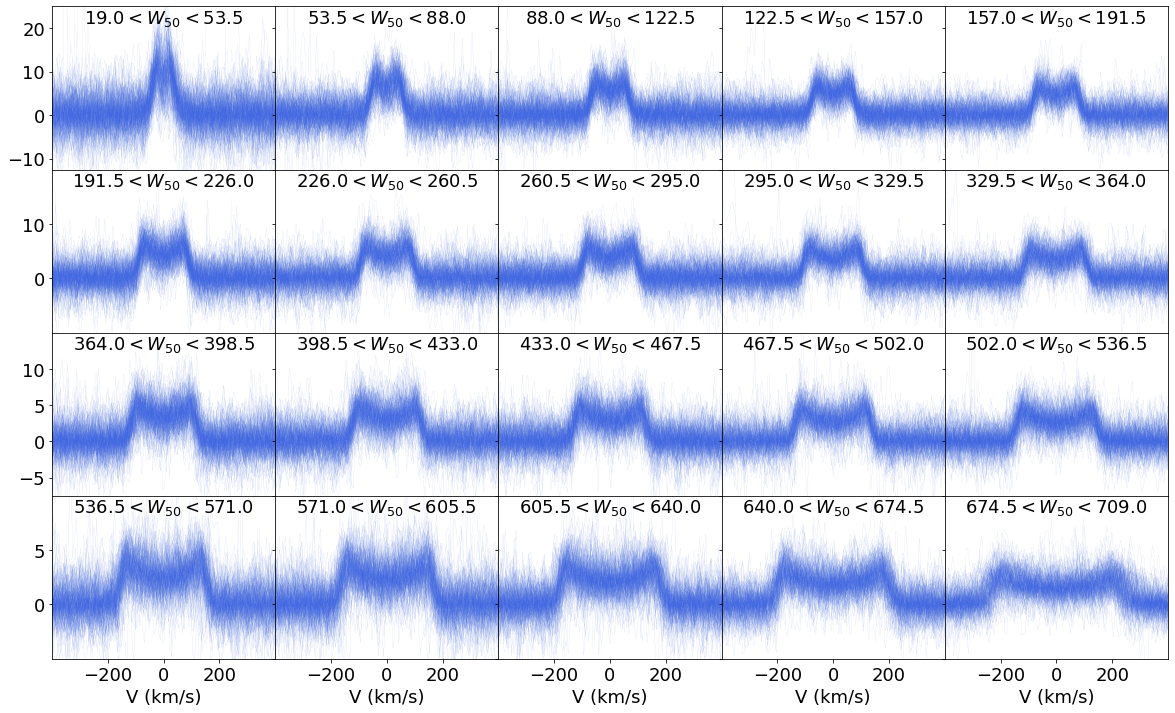}
    \caption{\HI integrated spectra of galaxies in the ALFALFA parent sample, divided into 20 equal-number bins by $W_{50}$ (km/s). In each panel, the flux is normalised by the total integrated flux, and the corresponding $W_{50}$ range is indicated.}
    \label{fig:ALFALFA_sample}
\end{figure*}

We show the \HI integrated spectra of 4844 galaxies in the ALFALFA parent sample, divided into 20 equal-number bins by $W_{50}$, as shown in Figure~\ref{fig:ALFALFA_sample}. The spectra are normalised by their total flux to highlight the characteristic line shapes. A prominent feature emerging from these stacked spectra is the prevalence of the double-horned profile, a typical signature of rotation-dominated \HI discs. This demonstrates the consistency of our sample selection criteria, which preferentially include galaxies with extended, regularly rotating gas discs.

\section{Definition of \HI mass in simulations}
\label{appendix: calibration}
In observations, the total \HI mass is measured through the telescope aperture, which typically extends well beyond the `true' \HI radius, defined as the radius where the \HI column density drops to $1\,M_\odot\,\mathrm{pc}^{-2}$. As a result, the observed $M_\mathrm{HI}$ generally includes gas at much lower surface densities beyond this `true' \HI radius.

In simulations, simply taking the \HI mass of the entire subhalo leads to a systematic offset in the $M_\mathrm{HI}$–$R_\mathrm{HI}$ relation compared to the empirical relation of~\citet{2016MNRAS.460.2143W}, whereas restricting the \HI mass strictly to within this `true' \HI radius underestimates the observationally accessible \HI content, as it ignores the extended low-column-density component.

\begin{figure}
    \centering
    \includegraphics[width=\hsize]{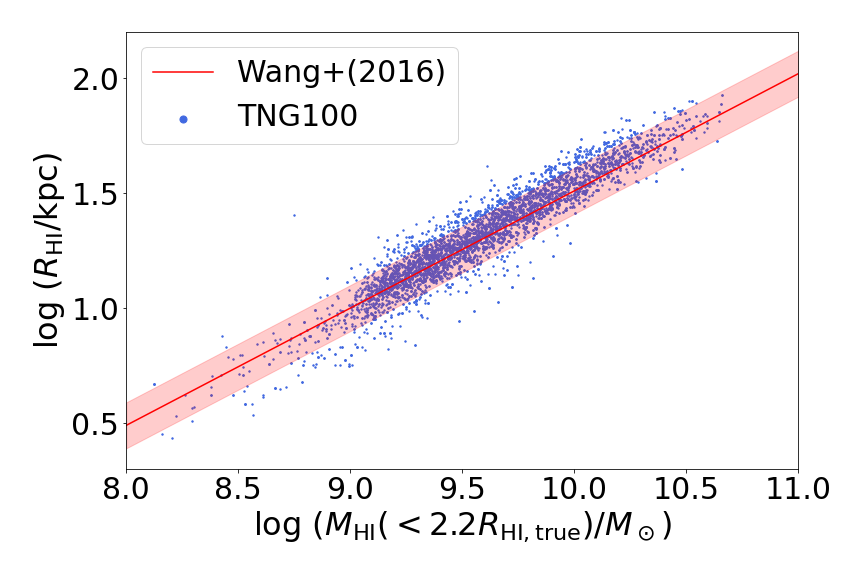}
    \caption{The correlation between the `true' \HI radius and the \HI mass within 2.2 times the `true' \HI radius. The TNG100 galaxies are displayed in blue dots, and the calibration of~\citet{2016MNRAS.460.2143W} is shown in red.}
    \label{fig:mhi-rhi-tng}
\end{figure}

We tested several alternative definitions and find that adopting the \HI mass enclosed within $2.2$ times the `true' \HI radius provides a good compromise: the resulting $M_\mathrm{HI}$–$R_\mathrm{HI}$ relation for TNG100 galaxies closely follows the empirical relation of~\citet{2016MNRAS.460.2143W} with only small scatter, as shown in Figure~\ref{fig:mhi-rhi-tng}. We therefore adopt this definition as the \HI mass and infer the observed-style $R_\mathrm{HI}$ from it.
\end{appendix}

\end{document}